# Building New Clubhouses: Bridging Refugee and Migrant Women into Technology Design and Production by Leveraging Assets

SONALI HEDDITCH, University of Queensland, Australia
DHAVAL VYAS, University of Queensland, Australia

While HCI scholars have examined how e-textiles serve to bridge the gender divide, there is little research into refugee, asylum seeker and low socioeconomic migrant women (WRAMs) and e-textiles. This paper presents the results of a series of two community-led participatory design workshops to study the factors that enable these women, who face intersecting barriers, to engage in STEM oriented making activities. Our findings examine 1) deficit discourse and strengths-based narratives; 2) bridging STEM skills into a culturally safe and tailored learning environment; 3) bridging commitment through commercial viability; and 4) the benefits of organizational partnering to bridge skills and diverse communities.  This paper makes three contributions. First, we offer a strengths-based counter narrative on the abilities, assets and motivations of WRAMs to engage in makerspaces, particularly STEM skills. Second, we offer a discussion on the implications of racial capitalism and internalized bias which limits resources, research and practice with WRAMs and consequently, technological design and production. Third, we extend the work of Buechley and contribute five strategies to bridge WRAMs into STEM oriented makerspace activities to build a "new clubhouse". We discuss the vital role researchers, technologists, makerspaces and financiers must play in supporting these new clubhouses to facilitate strengths-based narratives, harnessing and amplifying skills-based assets, in order to diversify who shapes technology and thus what is shaped.

CCS Concepts: • **Computer human interaction → Makerspace**

**KEYWORDS:** ACM proceedings, text tagging

## 1 INTRODUCTION

The International Organization for Migration (IOM) and the United Nations High Commissioner for Refugees (UNHCR) reported, as of late 2020, there have been over 317 million internal and external migrants, of which 79.5 million are forcibly displaced (e.g. refugees, internally displaced populations (IDPs)), with voluntary and forced mobility continuing to rise globally [94]. In 2020, the Australian Bureau of Statistics reported that half of Australia's population is either first or second-generation migrants [123]. Thirty percent of Australia's population are first generation migrants, equating to one of the highest amounts of foreign-born residents in the world, as well as one of the highest immigration rates in the world. The inner suburbs of metropolitan cities have a population between 32 – 40% who were born overseas, with the urban fringes often being home to 40% migrants and refugees. During the period 2011 – 2022, Australia resettled between 6000 – 22,000 refugees on humanitarian visas each year [20].

In this paper, the focus is on the most marginalized refugee and migrant women in Australia. It is important to acknowledge the difference between refugees, who are forcibly displaced, and migrants, who arrive in Australia of their own free will, with a significant proportion of migrants arriving with skills, resources, education and employment pathways available. However, Australian government data records show that migrant women are 7 per cent less likely to be employed than those born in Australia, and migrant and refugee women have a significantly lower rate of workforce participation compared to migrant and refugee men (47.3 per cent and 69.5 per cent respectively) [7]. Migrant and refugee women are overrepresented in non-professional care occupations, including child carers, personal care assistants, and aged and disability carers [7]. It has been consistently documented that migrant women from a non-English speaking background are far more likely to be engaged in secondary sector occupations, which pay low wages, have minimum holiday entitlements and no promotion ladder, as opposed to primary sector employment that provides high pay and favourable working conditions. In a 2020 study by Haque, it was found that non-English speaking migrant women were 23% less likely to be employed in primary sector employment, even for those with tertiary or vocational training and experience, with those women experiencing skill wastage and atrophy and exploitation, forcing migrants to accept lower level employment [46]. This paper focuses on refugee and migrant women (WRAMs) who possess a variety of assets that could be leveraged to make positive contributions towards Australia's workforce and innovation environment. However, these WRAMs face barriers in pursuing formal vocational training and skills development and consequently employment pathways. The refugee and migrant women involved in this study all reported facing challenges in successfully being able to navigate formal education such as vocational and technical training or tertiary studies, due to various barriers such as limited language literacy in their native language and/or English, constrained childhood education due to living in a crisis situation or experiencing discriminatory cultural norms, limited digital literacy, a lack of resourcing, a lack of social capital, or not having their education from their home country being recognized by Australia. The WRAMs in this study, despite possessing significant assets and strengths, have faced a lifetime of intersecting social, cultural and economic constraints, which have completely impeded any opportunities for any form of formal education or pursuit of primary sector pathways, including in STEM related fields.

The merits of utilising community based makerspaces to foster open innovation in an informal learning environment is widely recognised in Human Computer Interaction (HCI) literature [44], with the value of self-directed, experiential learning and the resulting self-empowerment extensively explored [34, 45, 58, 70, 95]. There has also been a proliferation of makerspaces in schools and universities to build Science, Technology, Engineering and Mathematics (STEM) skills and design capabilities for students [30, 34, 44, 58, 70]. In community settings, the values of democratic participation and low barriers to entry are core tenants of makerspaces [41, 105].

However, there has been an emerging body of work questioning makerspace inclusivity for novices and people of diverse backgrounds [26, 74, 80]. Skills related to traditional making (for example, wood and metal work, electronics) and technological making activities are often perceived as the domain of men [36, 38, 51, 89].

HCI scholars have studied women's access and utilization of makerspaces, examining how female makers persist in co-ed makerspaces [67, 81, 100, 109]. Over the last decade, HCI scholars have also researched the dynamics of feminist hackerspaces across the Pacific North West of the United States [36, 38, 89], Canada [35], and Sweden [65], as well as how women's traditional crafting activities have been examined and valued as a maker activity [37, 75, 85, 90, 115], with Buechley and Rosner leading insights on how such activities can be applied in a STEM context [92] and merged with electronics and computing [10–13, 15]. Recent research also examines makerspaces established for particular demographics who traditionally have limited access to formal and informal learning environments, such as people from a low socio-economic background [24, 106, 117], refugees [4, 5, 16, 27, 34, 102] as well as in developing country [32] and humanitarian contexts [21, 102]. However, it has also been observed that social and educational opportunities that do exist for the most marginalised WRAMs are frequently limited to craft-based skills, which can be attributed to global social norms which class craft as women's work [51, 54, 114]. As documented by Haque, employment opportunities are frequently gendered, precarious and exploitative, in sectors such as retail, hospitality and care and entrepreneurship pathways pursued are of the micro-entrepreneurship, craft dominated pathway.

In HCI, scholars have observed "identity-focused research tends to analyze one facet of identity at a time", with research on ethnicity and race lagging behind research on gender and socio-economic class [98]. There is a lacuna in the HCI field on intersectionality, which takes a more pluralistic perspective on the compounding impact of gender and other factors by which a woman is discriminated against (eg race, ethnicity, socio-economic status, disability, sexuality etc) [19, 23].

While HCI scholars such as Rosner, Buechley, Jones and others have sought to bridge women into electronics through the use of e-textiles [11–14, 63, 64, 92], there is a gap in the research focused on WRAMs. It is important to study this group specifically, because they face additional constraints to accessing makerspaces than white women [52–54]. This research aims to extend the work of Buechley and others into the domain of utilizing e-textiles with WRAMs and to offer lessons for engaging with this group.

We do this through two community-led participatory design workshops, introducing e-textile activities to a total of 15 WRAMs, along with a partner organization that offers craft-based activities to the women, and a feminist hackerspace. We examine four main findings, which have all been themed around types of 'bridging' required to enable the WRAMs to engage in STEM activities: 1) moving from a deficit discourse to a strengths-based narrative; 2) bridging STEM skills into a culturally safe and tailored learning environment; 3) bridging commitment through commercially viable production; and 4) the benefits of organizational partnering to bridge skills and diverse communities. This paper makes three contributions. First, we offer a strengths-based counter narrative on the abilities and motivations of WRAMs to engage in makerspaces, particularly STEM skills. Second, we offer a discussion on the implications of racial capitalism and internalized bias which limits resources, research and practice with WRAMs and consequently, technological design and production. Third, we extend the work of Buechley and contribute five strategies to bridge WRAMs into STEM oriented makerspace activities to build a "new clubhouse". We discuss the vital role researchers, technologists, makerspaces and financiers must play in supporting these new clubhouses to facilitate strengths-based narratives, harnessing and amplifying skills-based assets, in in order to diversify who shapes technology and thus what is shaped.

## 2 RELATED WORK

### 2.1 The Exclusivity of Technological Design and Production

In the 1980s, STS scholars began questioning the power dynamic that emerges with the differing access and use of technology between the sexes [68–70] and how technology can reinforce patriarchal power relations. Feminist STS scholars Donna Haraway, Evelyn Fox Keller and Sandra Harding challenged the idea that scientific knowledge is objective, with Keller [66] applying object/relations theory to demonstrate that science is 'masculinist', rather than human. Haraway questioned the 'situated knowledge', meaning the social context in which science is generated [47]. In the 1990s, standpoint theory was also integral to bringing intersectional perspectives to the scientific research lens, arguing that marginal lives, including those of women, people of color, of different sexual orientations, and others lacking social and economic privilege, must be considered [48]. In addressing standpoint theory, Harding called into question the singular category 'woman' which disregards the important intersections of race, class, ability, sexuality etc. Harding drew on the concept of intersectionality as defined by Kimberle Crenshaw [23]. Despite these questions being raised for decades by STS scholars, as well as increasingly amongst the HCI community, most technology continues to be designed by those with the privilege to obtain the skills and resources to be engaged in technological design, with reliance on biased data that reflects the needs of privileged populations, leading to products, services and infrastructure designed predominantly for white men [97]. This issue is on the verge of crisis, as we move at a rapid pace towards the AI revolution, which relies on data with massive gaps which renders Artificial Intelligence (AI) racist and sexist [96].

Women of color [29, 31, 49, 49, 86], including WRAMs [1, 52, 54] are often ostracized from technological design and development. HCI researchers are seeking to disrupt the heteropatriarchal structures of technological production by working with minorities to include them in technological design and production[3, 4, 87, 106], including speculative design and futuring [49], but these efforts remain sparse and ineffective in creating any systemic change. One paper worth noting is that of Knowles et al, which examines the experience of older adults, and how conflating aging with accessibility issues inadvertently harms older adults [69]. Knowles et al argue that focusing on age-related limitations perpetuates negative stereotypes of aging and promotes ageism and results in 'othering' of older adults. Assumptions that older adults lack the ability to use digital technologies makes it harder to conceive of meaningful contributions they might make as co-designers of technology futures. Knowles et al make recommendations to change these norms, including shifting mindsets towards positive narratives on ageing, and enfolding older adult perspectives into user design based on other shared contextual factors, rather than tethered to age [69].

While not focussed on technological design or production, there is extensive CSCW research examining how different marginalised user groups adopt technologies and their experiences with technology, including the benefits and challenges that arise due to having marginalised identity/ies. Pei examines migrants and refugees navigating the digital divide at a literacy centre, acknowledging the benefits but also the costs [82]. Afnan et al examine the privacy concerns held by Muslim-American women and the unique risks they experience online due to their intersectional identities [2], elaborating how the different intersecting marginalized identities amplify the constraints but also enable unique perspectives and experiences. The paper offers recommendations on privacy protection strategies, drawing on the lessons from the participant's unique experiences. Some recent CSCW research looks at broader structural issues, such as racial capitalism, with creative professionals of colour being short changed for the value they produce, such as having their work appropriated and monetised by the mainstream through social media and online platforms [113].

### 2.2 Makerspaces as Democratic Sites of Technological Design and Production

Makerspaces are venues which offer the potential to democratize technological design and production [25, 99, 104, 106, 117], and in turn to catalyze systemic change. Making activities in makerspaces can include traditional work with wood and metal, and handicrafts such as sewing, as well as more technological making activities such as digital fabrication via 3D printers, laser cutters and CNC routers [52, 99, 104]. There is a healthy level of critique of makerspaces not achieving their democratic goals, by reproducing heteropatriarchal structures and neglecting the involvement of marginalized communities [6, 26, 74, 79, 104].

There is also an emerging body of research studying makerspaces that cater to specific communities who have had limited access to education, such as people from a low socio-economic background [24, 106, 117], refugees [4, 34, 102] as well as in developing country [30] and humanitarian contexts [21]. HCI researchers have also examined feminist hackerspaces in the United States [36, 38, 88], Canada [35], and Sweden [65], and female makers in co-ed makerspaces [100]. Feminist hackerspaces have been established as an "interventionist response" to the dominant makerspace culture [121], with the focus on safety and shared values [36, 39, 88]. The researchers questioned the access for women from diverse backgrounds, observing that the majority of feminist hackerspace participants were white, well-educated, well-paid and predominantly from STEM fields [39]. The founders, cognizant of this situation, explained their visions to cater to ethnic and socioeconomic diversity, but failed to attract a diverse audience. One president of a feminist hackerspace stated:

> [A] lot of us are nice white ladies and we try not to be jerks about it and we really - try - we try to look at history and try not to do the same damn [thing...]. I would feel weird being all yes, we're an antiracist hackerspace when most of our members are white [38].

Beyond feminist hackerspaces in high income countries, there has been limited research at the intersection of HCI4D and Feminist HCI, which identifies the patriarchal structures that women in developing contexts are entrenched in, which constrains their engagement with technology [103].

HCI researchers have also examined more traditional female maker activity, such as makerspaces focused on arts, craft, sewing [59, 74, 115] and specific craft disciplines such as knitting [90] which see much more female involvement, including refugee and migrant women [53, 54]. Refugee and migrant women are rarely involved in makerspaces with tools for technological production [52–54]. Whelan states:

> 'Maker' and 'crafter' are acknowledged within maker discourse and research as separate identities. This is problematic as technological tinkering is the main site for the construction of a maker identity...bound up with performances of middle-class white masculinity... The maker movement's claims of universality, when coupled with a de facto focus on technological modes of production, allows for old paradigms of exclusion to be reproduced comparatively unchecked. [121]

Vossoughi et al offer a critique of culturally normative definitions of making and argue that "the ways making and equity are conceptualised can either restrict or expand the possibility that the growing maker movement will contribute to intellectually generative and liberatory educational experiences" for people of colour [114]. Vossoughi's article references an interview in *The Atlantic* entitled "All Immigrants Are Artists", and the seam-stressing conducted by female immigrants as a matter of survival. The article does not recast the immigrant as a maker to legitimise her capabilities, but examines the racialized and gendered hierarchies that lead to her everyday practices, casting her instead as an artist [114]. Hedditch's recent work examines makerspace practices that can hinder or enable access for under-resourced women of color who are novices to making, predominantly migrants and refugee women. She examines what makerspaces can do to facilitate an enabling environment appropriate to be inclusive of women who face intersecting oppressions [54]. However, there is also a strong body of literature on the importance of centring civil society organizations (CSOs) when working with marginalised groups, particularly CSOs that are owned and operated by people from the target community itself. In these instances, community-led design practices are considered best practice to ensure any research is positioned to meet the community's actual needs, and driven by community demand [22, 53, 71, 93]. Barajas-Lopez considers the value given to different materials utilised in the maker movement, in the context of Indigenous making and working with natural materials such as clay, in order to connect with traditional practices and ancestors as well as drive future innovation [8]. Nakamura examines the history of Navajo women being employed in electronic circuit manufacturing on a reservation, and the marketing that occurred to present these activities as connected to traditional practices such as rug weaving and silversmithing in order to portray an empowerment story, rather than one acknowledging the exploitation and harms occurring as a result of the industry [77]. Nakamura reasons:

> Race and gender are themselves forms of flexible capital…. Latinas and Asian, African American, and, later, Indian women were all viewed as having "nimble fingers and passive personalities. American Indian women, as well as Mexican women working in maquiladoras, were described in much the same way as "Orientals": as ideal workers in the digital industries, because of their experience with fine crafting of jewelry and textiles.[77]

**2.3  Lowering Barriers to Entry**

Several HCI researchers have been engaged in designing systems to lower the entry barrier in order to support novices to learn hands-on maker skills [110]. There has been a relatively recent push to make these systems learning-centric, not just technology-centric[55, 56, 110–112], designing for inclusivity and accessibility, including prioritizing autodidactic methods [55, 110–112], but with rising recognition for the limitations to DIY approaches in resource constrained environments [53, 101].

A significant area of focus to lower barriers to entry has been electronics and simple coding. A common sight at any electronics bench in a makerspace is the Arduino, an inexpensive, open source, and relatively easy-to-use embedded computing platform, which can be applied in a large variety of contexts. It was rapidly adopted to teach interaction design, engineering, and computer science in schools and universities around the world, and is also prominent in makerspaces [13, 76]. Research has been done in school environments, including under resourced contexts, and with people living with disabilities[28] .

Some researchers have leveraged crafting and sewing activities for women to introduce electronics and digital tools, with notable success [9–11, 14, 91], including a focus on older adults [62] and at risk school girls [72]. One of the earliest studies by Daniela Rosner consisted of a series of participatory workshops to merge quilting and electronics which "highlights for HCI scholars that the worlds of hand-work and computing, or weaving and space travel, are not as separate as we might imagine them to be" [91]. In the research, core memory planes were used as quilt patches, along with conductive thread to sew the patches together. As part of the study, the researchers and participants also explored the underlying role of crafting and gendered-innovation in the development of electronics and computing, arguing for recognition of these legacies and the ongoing role of women to bring forth gendered innovations [91].

In 2006, HCI scholar Leah Buechley created the LilyPad Arduino, subsequently partnering with SparkFun Electronics to produce and sell the product commercially from 2007. The LilyPad

Arduino enables people to create their own electronic textiles or "e-textiles", consisting of a spool of conductive thread and a set of sewable electronic modules, including a sewable Arduino microcontroller, a temperature sensor, an accelerometer, and an RGB LED [12, 13]. E-textiles are articles of clothing, home furnishings or architectures that include embedded computational and electronic elements [11]. The LilyPad Arduino enables e-textile making to be accessible for novices [11]. E-textile activities included light up and audio-emitting stuffed toys, fashion and home wares.

In 2010, Buechley studied the sales and adoption of the LilyPad Arduino over a 2 year period, in comparison to Arduinos and found the former to have much higher take up among women (though still in a minority) compared to the latter. People who purchased Arduinos, were largely male (78%), with 9% female (and the remainder unknown). In contrast, 57% of LilyPad customers were male and 35% were female (with the remainder unknown). In studying projects undertaken, Buechley found 86% of Arduino projects were done by males and 2% by females. In contrast 25% of LilyPad projects were done by males and 65% by females [13].

Buechley states that "the LilyPad community ... confounds gender stereotypes and demographic patterns in electrical engineering and computer science—both overwhelmingly male dominated fields" [13]. She concludes:

> Our experience suggests a different approach, one we call Building New Clubhouses. Instead of trying to fit people into existing engineering cultures, it may be more constructive to try to spark and support new cultures, to build new clubhouses. Our experiences have led us to believe that the problem is not so much that communities are prejudiced or exclusive but that they're limited in breadth—both intellectually and culturally [13].

The aim of this research is to explore what is required to "build a new clubhouse" serving intersectional women, particularly WRAMs. Online searches and engagement with the makerspace community indicate that there are no makerspaces identified in Australia or internationally that achieve engagement of women from intersectional backgrounds, varying in terms of race, ethnicity, religion, educational level, socio economic status, disability, sexuality etc in the full spectrum of makerspace activities, rather than niched activities, usually craft. This research will draw on lessons to explore factors that could contribute to successful engagement of intersectional women in the maker movement as a gateway to involvement in technological design and production.

Some early work has been done to attempt to build new clubhouses. Desportes et al studied the design and development of a makerspace located within a predominantly Black neighbourhood in the United States, operated by a social justice organization [25], with findings centred on strengths and assets based approaches to building community and individual capacities. Carucci and Toyama established a temporary makerspace in an aged care home, to assess if the makerspace could improve well-being for residents [17]. This research aims to extend this preliminary work. By upskilling people of all backgrounds, with intersectional identities and different levels of education and resourcing into technological design and production, the innovations may be transformative.

### 2.4 Community-Led Participatory Design

There is a strong body of literature recognizing that universalist design privileges dominant cultures. Participation is a central concept in HCI in seeking to embrace democratic principles and involvement of "users" in design [3,46,50,69,83], often considered a sound method to engage with marginalized and underserved populations [50]. It has become the norm to involve participants in design, usually through workshops, for once off involvement, or perhaps a short series of events, to receive user group input. However, in participatory design, the lead designer is still the external expert, often not from the community of 'users', particularly when the design is for those who are intersectionally disadvantaged or multiply burdened under the matrix of domination. Critical scholarship questions who gets to participate and on what terms, asking what constitutes meaningful participation [19,22,83]. Scholars have sought to deconstruct the privilege of the participatory design approach within HCI and re-centre the focus of design on individuals who are historically underserved [50]. Vines et al. question how participation in HCI should be configured, challenging designers to consider the timescale of involvement (from short events to project based ongoing involvement), who initiates the design and nature of participation (with it almost always being the designer) and who controls the process, with participation masking the agency, expertise and agendas of the researchers or designers leading or facilitating the participatory process [83]. Bardzell postulates that feminism in HCI can contribute to an action-based design agenda, integrating critical thinking into all stages of the design process including user research, prototyping and evaluation [6]. Applying standpoint theory, Bardzell suggests that the introduction of new user research, the "marginal user" will occur if the alternative epistemology is privileged.

Duarte et al. examine participatory design and participatory research in a case study in Germany with young forced migrants to develop a technological application to support them on arrival and settlement [22]. These marginal users are centred in the process, with important findings on the advantages and disadvantages of participatory design and research approaches with a forced migrants who are struggling with trust in their new society and language. Findings centre on creating a safe space for an iterative process, with challenges remaining to engage the youth in a formal education setting and overcoming language barriers. The researchers and authors of Duarte's paper remain centred as the facilitators of the process. In a paper on "An Intersectional Approach to Designing in the Margins' Erete et al. contend that: "Traditional research design methods (e.g., interviews, surveys), and even those participatory in nature, at times do not match the needs of our participants, leading to questions regarding the effectiveness of these methods among certain populations" [23]. Harrington espouses a more collaborative approach, seeking to elevate the voices of particular communities, while also addressing issues of power and positionality [50].

Constanza-Chock calls for 'design justice' [19]. At its core, design justice demands that design is led by the marginalised communities for which the design is intended. By positioning the community as the design leads, Costanza-Chock argues that this will dismantle structural inequality and address all aspects of the needs of intersectional identities. The design justice framework builds off black feminist thought and seeks to decolonize design, presenting 10 principles to achieve design justice, which can be reviewed at www.designjustice.org. While the workshops in this study consisted of only two days, rather than an extended engagement, the principles of elevating the user and positioning WRAMs as design leads were aimed for during the conceptualisation and planning of the engagement itself as well as the workshop engagement. Thus, we define the method as "a community-led participatory approach" to working with WRAMs.

Acknowledging the criticism of interventions based on needs based frameworks, where needs are met by expert outsides, Pei considers Assets-Based Community Development (ABCD) which starts with what is already present in the community, including the capacity of the community themselves [83]. Pei proposes a framework to evaluate the potential of an intervention to deliver sustainable impact in a resource constrained setting, as represented in the figure below. The horizontal axis represents novelty, with interventions low in novelty more likely to deliver sustainable impact. The vertical axis represents asset utilization, with interventions high in asset utilization more likely to deliver sustainable impact [83].

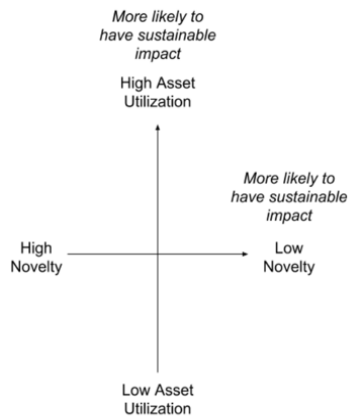

Figure 1: A framework for sustainable impact [83]

Pei distinguishes between invention of new technology (high in novelty), adoption of new technology (high in asset utilization and medium in novelty) and amplification of technology already used in the community (low in novelty), with the lowest in novelty being the most sustainable [83].

## 3 STUDY

### 3.1 Study Objectives

The objectives of this study were to:
- explore methods to bridge WRAMs active in a craft-based organization into non-gendered maker skills;
- assess the outcomes of the process and activities;
- gather lessons for craft-oriented organizations, makerspaces, feminist hackerspaces, HCI researchers, designers and their collaborators more broadly who are seeking mechanisms to disrupt bias in technological production.

### 3.2 The First Bridge: Bringing Together Two Organizations

In order to undertake the study, we commenced by establishing partnerships with two distinct organizations, for the purpose of bridging WRAMs to a broader range of maker activities.

The first organization, CraftPlace, has been operating for 10 years, aimed at WRAMs. These women are known as 'sisters'. CraftPlace itself is predominantly staffed by women who arrived in Australia as asylum seekers, refugees and migrants. Overall, the staff tend to have a stronger educational background and work experience than the sisters, with more confidence to navigate Australian society. For some staff, they started their journey as sisters at CraftPlace, and as their capacity for teaching, administrative or management roles emerged, they were offered staff positions. The organization is well respected as a leading institution for their sisters' empowerment. This is attributed to being managed by women with lived experience, as well as a broader mandate than similar organizations that focus only on craft, community building and English language acquisition. (Other craft organizations considered by the first author were run by church groups and community organizations, with 'white' management teams who had been in Australia for multiple generations). CraftPlace's distinction comes from income generation pathways, primarily micro entrepreneurship and more recently, casual employment. The nature of the work offered has been very gendered, focused on craft-based activities, such as sewing, crochet, jewellery making, candle making, preparation of bees wax wraps, screen-printing and cooking of saleable condiments, such as chutneys and jams. Women involved sell the products they make predominantly through market stalls (micro-entrepreneurship) or gain employment in the makerspace to produce what are known as CraftPlace 'label' products. CraftPlace sells these label products in a retail venue in front of the makerspace, as well as wholesales to corporate partners, with rising success, due to the social impact being generated for the female asylum seeker, refugee and migrant community. CraftPlace also facilitates employment for the sisters at external employers. The sisters generally gain employment in sewing, packaging, hospitality and retail sectors.

The second organization, DigitalCastle, had been operating for 3 years as a feminist hackerspace, offering predominantly STEM based activities, leveraging creative technology, such as e-textiles, digital fabrication, electronics, coding and music production. DigitalCastle's is a relatively small operation, with one founder who conducts workshops herself and occasionally outsources to her network. It is currently the only feminist hackerspace in Australia known to the founder and authors. The participants are mostly women and marginalized sexualities and genders (from the LGTBQIA+ community) who are involved in the art and digital technologies scene, professionally or as enthusiastic hobbyists. When the first author commenced engaging with DigitalCastle, they had yet to engage with any women of color, with no experience working with WRAMs.

The first author approached CraftPlace to introduce the idea of bridging pathways into STEM for the refugee and migrant women initially in September 2019. The concept was met with a desire to support the first author in her research, but an overall feeling that the sisters themselves would not be interested. Several options were put forward, including DIY biology[33] projects such as kombucha or mycelium leather production, e-waste activities such as computer disassembly and reassembly or e-waste jewellery making with a soldering component [52, 68, 116, 117], utilizing simple maker kits on the market with an electronics focus [63, 64, 84], basic drag and drop coding for 3D printing and digital craft [40, 78], and other citizen science type activities, but none were found to particularly resonate with the CraftPlace staff as having potential appeal to their sisters.

In November 2021, the first author met the founder of DigitalCastle. During this meeting, they discussed that DigitalCastle had yet to attract any women of color, including refugees and migrants to their venue. They discussed the marketing strategies employed, but like other feminist hackerspaces in other contexts globally[36, 38, 88], had struggled to understand how to successfully bridge to this audience. The founder was open to the concept of partnering with an organization such as CraftPlace to bridge this gap.

In the first half of 2022, the first author secured a commitment from CraftPlace to trial two days of community-led participatory design workshops on e-textiles. Drawing on an assets-based approach, it was believed that this would be the most appropriate pathway to enable women, who were familiar with sewing, to try something new, by learning about circuits and electronics. DigitalCastle also made a commitment during this period to join the collaboration, to teach e-textiles, and to gain experience working with the WRAM community. This was not a 'simple' or necessarily sustainable solution, as the organizations are based in two different cities, 2.5 hours apart by flight.

### 3.3 The Sisters

Due to concerns from CraftPlace about low attendance, sisters were recruited through a gentle 'push' method for Day 1, where the sisters were put in a position where it would require minimal additional effort for them to join our workshops. Six sisters, from Vietnam, Indonesia, China, India, Hong Kong and the Philippines, were invited to join the Day 1 workshop as a final day in their sewing course. These sisters will be referred to as 'Group A'. While not a compulsory requirement, the workshop was scheduled at the same day and time as the previous weekly sewing course sessions, so offered as an extension activity. Further, on the day of the workshop, interviews were being conducted to become a casual employee at CraftPlace. As graduates of the CraftPlace sewing course, these sisters were eligible to be interviewed for employment to produce CraftPlace label products. For all of these sisters, it was their first opportunity to secure employment in Australia, and thus, they would be motivated to come to CraftPlace and thus likely to attend Day 1 of the workshop. We designed the workshops so that Group A sisters could return on Day 2 to extend their learning, but there would be no compelling reason to return such as the interview, and thus expectations were low on there being returnees. In fact, only 1 sister, at time of recruitment, indicated she may return for Day 2. However, in practice, all 6 sisters indicated the desire to return after participating on Day 1, and 4 did return for Day 2. The other 2 were unable to due to previous commitments.

The Group B sisters were recruited through a 'pull' method for Day 2, where active recruitment was required and sisters would need to elect to come of their own volition. Posters were produced and placed around the CraftPlace venue advertising this special workshop, outside the norm of regular programming. Low numbers were expected, with 5 women signing up. However, 9 women attended, ranging in ages from early 20s to late 60s, from Turkey, India, China, Cameroon, Congo, Venezuela, Hong Kong and Vietnam.

All sisters understood that this was for an independent research study and signed consent forms regarding their participation and use of data.

### 3.4 Positionality

The first author is the daughter of a Sri Lankan migrant, and her mother did not complete her high school education due to cultural norms and expectation for her to take care of the family. The first author's father, while white, was also not able to complete his high school education, entering the world of full-time work at the age of 14. It was due to their own circumstances that the parents of the first author emphasized the importance of education for the first author, but she still experienced cultural norms regarding gendered roles. The first author witnessed her own mother struggle to pursue any pathways back to skill building that could lead to meaningful employment beyond gendered activities such as 'family day care' in the home, retail and cleaning. Unfortunately, avenues remain very constrained for female refugees and migrants in Australia who missed out on an education and continue to miss out due to a lack of networks, resources, venues and discriminatory social and cultural norms. The first author perceives makerspaces as potential sites of disruption due to her own shaping.

### 3.5 Conceptualization, Planning, Preparatory Interviews, Workshops, Debriefing Interviews

Table 1 below documents a two staged process, with Stage 1 engagements occurring with the partner organisations, (CraftPlace and DigitalCastle) from September 2019 to August 2022, for the purpose of conceptualizing and planning the workshops, which required several iterations to ensure an approach that was community-led by CraftPlace to meet the needs and interests of the sisters and CraftPlace itself, but also within the capabilities and interests of DigitalCastle and in line with the studies' research objectives. In Table 1, Stage 2 captures the workshops that were conducted on Day 1 with Group A (6 sisters), with Day 2 extending Day 1 learnings for 4 of the 6 sisters. On Day 2, Group B, consisting of 9 sisters, attended to undertake the same workshop content offered to Group A on Day 1. It also documents the post workshop debriefings and analysis.

Table 1: Engagements with Stakeholders

| Stage | Timing | Participants | Nature of engagement | Length |
|---|---|---|---|---|
| **Stage 1** | September 2019 – June 2022 | CraftPlace staff, including the CEO, Head of the Empowerment Hub, and Head of Production | Discussions on the concept of trialing STEM skills with makerspace participants, progressing from conceptual to practical plans | Approximately 6 x 15 minute discussions on concepts, and 2 x 30 minutes on planning, plus emails with details, 3 hours total including emailing. |
| | November 2021 – June 2022 | DigitalCastle founder | Discussion on the concept of reaching refugee and migrant women as participants, through to practical planning | Approximately 5 x 30 minute discussions, plus planning emails, total 3 hours including emails |
| | July – August 2022 | CraftPlace Head of Production | Preparatory interviews, exchanges on practical planning, briefings on Sister participants | 3 hours |
| | July – August 2022 | DigitalCastle Founder | Preparatory interviews, exchanges on practical planning of methods and activities, workshop formats, materials, presentation format | 3 hours |
| **Stage 2** | September 8, 2022 | CraftPlace and DigitalCastle Staff<br>6 refugee and migrant women (Group A) | E-textiles Workshop (Day 1 for Group A)<br>Individual interviews with Group A women, as well collective discussion and reflections | 6 hours |
| | September 9, 2022 | CraftPlace and DigitalCastle Staff<br>6 refugee and migrant women (Group A)<br>9 refugee and migrant women (Group B) | E-textiles Workshop (Day 2 for Group A)<br>E-textiles workshop (Day 1, Group B)<br>Individual interviews with Group B women, as well as collective discussion and reflections<br>Further discussions with Group A women on Day 2 learnings | 8 hours |
| | September 13 | Follow up interviews with workshop participants | Individual interviews with 2 participants from Group A | 2 hours |
| | September 15 | Follow up joint interview with DigitalCastle founder and CraftPlace Head of Production and Head of Empowerment Hub | Discussion on lessons learned from workshops | 1 hour |
| | TOTAL | | | 29 hours |

### 3.6 Workshop agenda

The workshop agenda was planned by the researchers in close coordination with DigitalCastle, and CraftPlace, with the sewing skill assets of the WRAMs centered and elevated through CraftPlace guidance. We privileged visual and tactile content to support comprehension by those with limited English, as well as visual participatory contributions by the participants. The workshops were facilitated by the founder of DigitalCastle and the lead author, supported by the CraftPlace Head of Production to ensure institutional knowledge transfer. In addition, one or two CraftPlace volunteers attended at various stages throughout the day to support comprehension by the sisters with limited English.

Table 2: Workshop Agenda (Day 1 and Day 2)

DAY 1 Agenda (Group A and Group B)

| Time | Activity |
|---|---|
| 9:30 | Introductions through a drawing activity. Drawing was selected as a method of expression in order to set a calm tone, to help facilitate communication for those with limited English, to understand constraints and opportunities and to get the creative juices flowing. Participants then presented the drawing to the group. |
| 10:15 | Presentation on electronics and e-textiles by DigitalCastle founder and first author |
| 10:45 | Design. Participants asked to choose an embroidery design based on their drawing from the introductory activity, and draw it on fabric. Assistance was given to map out the placement of LED lights and how to sew the circuit, without positive and negative lines touching, and testing LEDs to ensure they were working. |
| 11:15 | Execution of design. Sewing of circuit and decoration of canvas through mixed media such as felt, paint, coloring pencils and embroidery. One on one interviews also conducted during this time. |
| 12:30: | Lunch break over communal food, during which informal discussions were shared on the learnings. |
| 13:15 | Execution of design. Continuation of circuit sewing and decoration of canvas. One on one interviews also conducted during this session. |
| 14:15 | Presentation of work to the group, followed by an ideation session. Discussion of how skills may be utilized in future, and any requirements for future learning. |

DAY 2 Agenda (Group A only)

| Time | Activity |
|---|---|
| 9:30 | Extension into using RGB LED Lights. Presentation on activity, materials, and activity. Product discussion. |
| 10:00 | Prototyping for product development. |
| 14:00 | Feedback session on needs for product development |

### 3.7 Data Collection and Analysis

Data was collected through a mixed methods approach throughout both stages of the research process, including author Stage 1 notes on initial discussions with the partner organisations to plan for the research project and to establish the partnership, as well as to plan the community-led participatory workshops. During the preparatory interviews and Stage 2 workshops, extensive notes were also maintained to capture the data from the semi-structured interviews with staff from both organizations and sisters, as well as audio recordings to cross-reference with the notes. Notes were also taken to document the ethnographic observations made while sisters were making, along with notes on accompanying informal conversations. Photography, videography and cataloguing of objects also occurred. The drawings undertaken by the sisters during the introductory session of the workshops were kept by the researchers, along with separate commentary from observations and interviews.

The data was coded by thematic analysis [18], against the two research stages, as the Stage 1 process in itself revealed many important lessons on community-led participatory design and assets based approaches. In terms of Stage 2, the data was coded against the three research objectives stated in 3.1 above. For both Stage 1 and Stage 2 sets of data, an iterative process was used to refine codes within our data, with themes collated across four main findings across both stages.

## 4 FINDINGS

The four main findings have all been themed around types of 'bridging' required to enable the WRAMs to engage in STEM activities. The findings are summarized in 4.5, through the lens of Assets and Constraints, as identified by CraftPlace during Stage 1 and Stage 2, and separately by the WRAMs themselves during Stage 2.

### 4.1 Moving from a Deficit Discourse to a Strengths-Based Narrative, Bridged by the WRAMs Words and Actions

During the Stage 1 conceptualization of the project with CraftPlace, as documented in the methods section 3.2, and in selection of participants in 3.3, the selection of the research activity and workshop format was guided by the staff at CraftPlace, drawing on an assets-based approach. The staff at CraftPlace are themselves often migrants who have experienced similar constraints to the sisters, but not to the same degree. In employing principles of community-led design [22, 27, 50, 53], the concerns of the staff at CraftPlace were centered. As a result of these concerns, e-textile activities were selected by CraftPlace, due to it being perceived as the closest activity to the regular sewing activities undertaken and therefore harnessing existing assets. However, many concerns remained regarding the inclination of sisters to be involved or to actively participate. Key concerns included that sisters:

- would not have any interest in any activities involving STEM activities including electronics, and thus it would be difficult to secure attendance of any sisters;
- may be interested, but also feel constrained from participating due to intersecting social and cultural norms regarding gender roles in their culture, with a conscious or unconscious bias against engaging in STEM related activities, or being restricted by male family members from participating;
- may be interested, but would lack confidence to engage in the workshops due to a lack of experience with anything related to STEM;
- would not be interested due to no obvious income stream or benefits from engaging;

- if presented with the equipment and materials to engage with e-textiles, would feel overwhelmed and thus unwilling to engage, given their lack of experience or exposure to such activities;
- would not be able to comprehend the activities and thus not capable of following the instructions;
- would not be capable of certain activities, particularly any coding, no matter how simple and intuitive.

In response to these doubts, it was agreed during the planning sessions that:

- CraftPlace would recruit attendees through both push and pull methods as outlined in section 3.3, with no guarantees they could secure attendance;
- E-textile activities would remain simple. No drag and drop programming would be included as an extension activity.

In the lead up to the workshops, including the day before, the lead staff member at CraftPlace warned the first author that it would be unlikely that anyone who attended on Day 1 would return for Day 2. She was also concerned that they would have very few new women turn up on Day 2 for the second round of the workshop, due to the lack of other compelling reasons to attend.

However, it was discovered by the staff at CraftPlace, and thus the authors, that their concern regarding no shows, a lack of repeat attendees and expected lack of attendees on Day 2 were unfounded. The staff at CraftPlace were surprised by the number of returnees on Day 2 and fresh attendees that day. The CraftPlace staff members also observed that usually, sisters always break for lunch, and given lunch was also catered on both workshop days as an additional incentive for sisters to attend, she was expecting the usual behaviour of a long, social lunch break. However, she noted that she could see how much the sisters were enjoying the workshops because they weren't coming out for lunch either day, and wanted to continue with the e-textile activities during the lunch break. The CraftPlace staff member stated to the first author:

> You've proved your hypothesis that participants here would like to be offered this type of program.

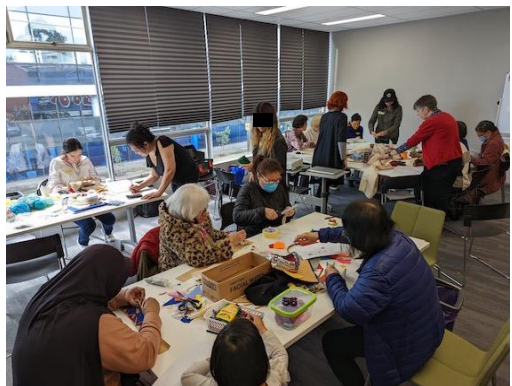

Fig. 1. A full house on Day 2

As a warm-up activity, in order to establish trust, but also to simultaneously understand the background of the women involved, sisters were asked to draw, rather than write (given English was a new language for most participants, and also to get the creative juices flowing), a picture that represents them. In seeking to understand their identity and intersectional and constraining aspects, as well as assets-based positioning, the women were also asked to describe what boxed them in, and also when they do not feel boxed in or constrained.

As sisters presented each of their drawings back to the group, common themes emerged for both Group A and Group B:

- Most sisters did not feel boxed in, but rather talked about their excitement to be in Australia, their gratitude to CraftPlace for creating a sense of belonging and a place to learn and connect, and their excitement to build skills and pursue new opportunities.
- For the sisters that did feel boxed in, the common constraint experienced was not being able to utilize their formal skills in Australia, with their qualifications not being recognized or employers declining their applications due to a lack of experience in Australia. A discussion was held around the impossibility of finding venues to gain experience in Australia to create pathways into employment aligned with their skill set. However, even for these sisters, they discussed the excitement for gaining new skills and pursuing new pathways and credited CraftPlace as the venue that made this possible. Sisters generally seemed genuine in having a growth mindset and determination to make the most of opportunities presented.

In interviewing sisters about their previous experiences in leveraging STEM skills or technology, they all revealed that the workshop activity was completely new (with the exception of one Vietnamese woman mentioned below). None of the sisters said they found it daunting or difficult, and wanted to learn more. They explained that the subject matter, of learning technology, was exciting. One older sister from the Philippines stated: "I always enjoy learning new things. That is what life is all about. I want to keep learning, to keep growing, it makes me feel alive." Another Chinese sister stated "I am enjoying this so much. It's waking up my brain. I have so many ideas on things I could do with this, new things I could learn. I want to keep learning more and more."

One of the Group A participants revealed that although she had not worked in 20 years, her first job in Vietnam had been in electronics. She said she enjoyed re-engaging in electronics and realized she would like to pursue employment in electronics again. The staff at CraftPlace had been unaware that this sister had experience in electronics.

When sisters were asked if they would engage with more STEM based workshops at CraftPlace, they unanimously agreed they would like to continue these sorts of learning activities. As noted previously, it was initially expected that none or only perhaps one of the sisters from Group A would return for Day 2, but at the conclusion of Day 1, all six Group A sisters wanted to return to extend their learning, but two could not return due to previous commitments. When asked what could constrain their future engagement in learning STEM skills, the sister from Indonesia explained "The main issue is English. The hesitation isn't the technology, its language."

This sister went on to explain that when she has attended other vocational training centres, she cannot understand due to the fast pace of English and technical language used. She knows that at CraftPlace, she will be taught in a way to support her comprehension. She expressed a strong desire to be offered a variety of maker skills, showing particular interest in digital fabrication when examples were shown to her. She wanted her strengths and abilities to be realised, but to be engaged in a way that was sensitive to her language limitations.

Overall, there was a substantial gap between how CraftPlace staff conceptualised the assets, interests and abilities of the sisters, and their actual interests, assets and abilities. During Stage 2, it became evident that the social construction of the limitations of the sisters, which resulted in some Stage 1 deficit discourse as outlined in the bullets above, was in stark contrast to the strengths-based dialogue utilized by the sisters themselves during the workshops and interviews.

## 4.2 Bridging STEM Skills into a Culturally Safe and Tailored Learning Environment

One of the key assets to the success of the workshops was the ability to leverage a venue the CraftPlace makerspace, with which sisters were already familiar and comfortable. The importance of utilizing CraftSpace's own venue were emphasized by their staff during Stage 1, and consequently, issues regarding social and cultural safety that can often arise for the sisters when attending unknown venues were absent. Women who did not have childcare were able to bring their children. Lunch was provided to foster community (although as noted above, women did not partake for long due to their enthusiasm to engage with the activities), and familiar staff and community volunteers were present.

With the venue not being a concern, the key challenge was in preparing content for the workshop that would be relatable, relevant and delivered in a way that ensured comprehension for the sisters, at an appropriate pace. CraftPlace staff also held substantial knowledge assets on how to create a learning environment that meets the needs of sisters. Some of the key principles applied in designing the workshop agenda and content, were:

*4.2.1 Elevating Visual Content for Explanations and Inspiration, Including Photos and Videos.* In advance of the workshops, the first author prepared a powerpoint, which included as much pictorial imagery as possible to offer explanation, given the limited English literacy of sisters (Figure 2). However, another purpose was to also evoke excitement for the workshop and the possibilities of e-textiles, to ensure engagement by the sisters. In addition to photos, sourced from DigitalCastle and online, video content was also sourced to demonstrate e-textile products.

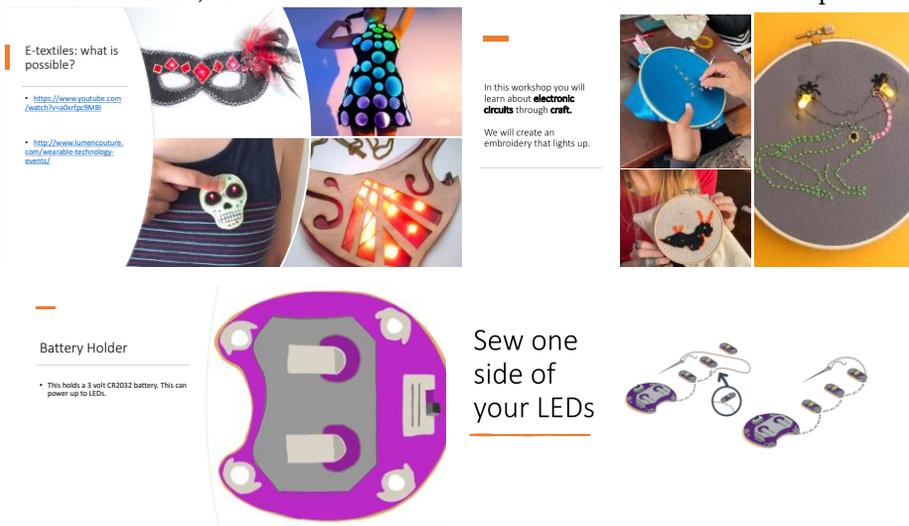

Fig. 2. Examples of some of the many highly visual slides used to facilitate learning

One difficulty encountered by the first author was sourcing any video content that did not feature white women, or English being spoken slowly. For this purpose, videos that relied on imagery to show products rather than any verbal explanations were prioritized. Videos with white American women were found, only delivered with fast paced English. In the end, some of these were still utilized due to the lacuna of content online featuring women of colour and e-textiles, but the first author switched off the audio and narrated herself, in order to support better comprehension.

Overall, the sisters responded well to the photo and video content. When making their own e-textiles, they kept referring back to the imagery for guidance. As will be discussed in 4.3, the photos and videos also fuelled product inspiration.

*4.2.2 Offering Tactile Opportunities to Engage with the Materials and Equipment.* In addition to the images, physical materials and equipment were discussed and passed around for the sisters to touch and engage with. Sisters were able to touch the battery holder, LEDs, conductive thread before progressing with their designs. Sisters were also provided with alligator clips to test their circuits and the LED lights as they progressed with laying out their design and sewing their circuits, to scaffold and reinforce learning (Figure 3).

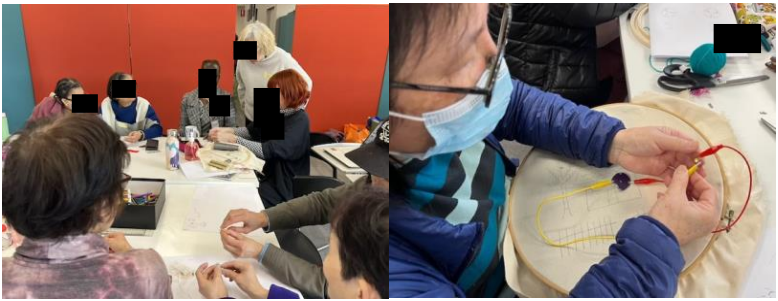

Fig. 3. Tactile opportunities to engage with the materials before commencing their design as well as for scaffolding

*4.2.3 Leveraging Staff, Community Volunteers and Peers to Enable Comprehension.* CraftPlace brings significant scaffolding to its learning environment, in the form of staff, community volunteers and peer support. During the process of sisters' designing their e-textile embroidery, planning their circuit, tracing it in and laying it out, substantial support was offered by the staff from CraftPlace, DigitalCastle, the first author, as well as community volunteers from CraftPlace. Peers also supported each other**.**

Even though there was plenty of consideration of how to best aid comprehension for the sisters, given their limited English and inexperience with e-textiles, the measures did not lead to immediate success. Mistakes were made, such as stitches being too long which reduced conductivity, one sister did not understand that she needed to sew from positive to positive (and negative to negative) to ensure a working circuit, and one sister began to sew her LEDs on the wrong side of the embroidery.  It was due to the ample level of staff, volunteer and peer support, that most issues were caught sufficiently early and addressed so each participant could leave the workshop with a functioning circuit, but errors remained, emphasizing the need for significant levels of scaffolding to aid comprehension.

**4.3 Bridging Commitment through Commercial Viability.**

At the end of Day 1, the sisters in both Groups A and B were offered the opportunity to present their work back to the group, as well as ideate on future e-textile product creation. To facilitate the ideation, a series of slides and a video were shown to stimulate thinking (Figure 4).

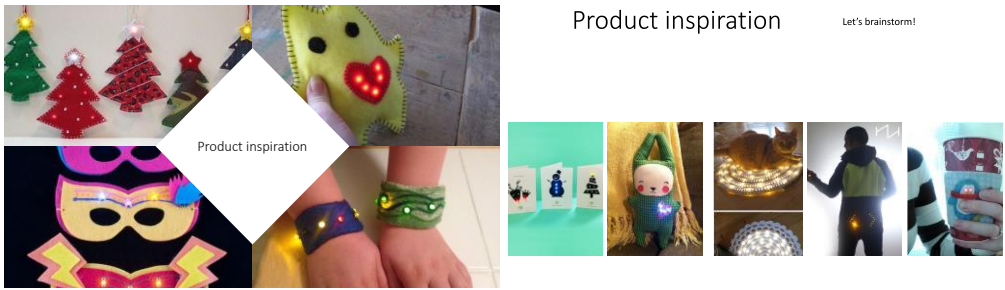

Fig. 4. Examples of product inspiration slides

Sisters demonstrated agile thinking about what is possible. For example, much discussion was held on how the electronic materials themselves need to be modified to better support novices and reduce potential for mistakes. Ideas included:

- having one flat edge and one round edge of the LED sequins to make it easier to distinguish the positive and negative ends, rather than both ends being rounded. This was particularly important for the older participants who had reduced eyesight, but all agreed this would be a helpful innovation;
- the need for soft LED sequins that can be sewn over with a sewing machine rather than needing to be attached by hand. After this suggestion, the sisters watched a video that demonstrated an industrial machine in action, but wanted to see sequins appropriate for domestic sewing machines;
- requesting handouts on where to source parts so they could make their own light up t-shirts, scrunchies and other products for their children and grandchildren, as well as for markets; and
- having access to making their own components to meet their needs and design ideas for products, due to dissatisfaction with the components on offer to work with at the workshop.

In contrast to the deficit discourse employed by CraftPlace staff in planning the workshop, and a focus on STEM skills-based constraints, the refugees and migrants demonstrated progressive, agile and innovative thinking around what would be possible with their new skills. When their agility was remarked upon by the first author, the refugees and migrants discussed how they were all very adaptive, given they have had to survive and adapt to new circumstances as part of their migration journey. It was collectively agreed that this agile mindset was a key strength, and that this strength should be harnessed to learn new skills and pursue new opportunities, such as STEM skills at CraftPlace.

A newly arrived woman from Group B noted how much this technology was needed in her home country and that she wished she could extend the skills back home:

> Where I'm from, there aren't many lights at night. It's not safe to walk in the dark. Can you imagine if everyone could make something to wear that could light up? Everyone would be so excited and want to do this. I wish we could.

This comment spurred a discussion about the safety potential of e-textiles, with ideas for various light up wearables for walking at night. A light up safety vest was discussed as a potential product, with CraftPlace and Digital Castle staff contributing ideas on how to make it suitable for bike riders.

For the four sisters from Group A who returned for Day 2, they extended their learning to RGB switches. They agreed collectively to make a crown. This was not their own prototype, but involved copying a design of the first author utilizing an RGB switch which had been shown as inspiration. However, they liked this product because it aligned with many of the ideas they had regarding the purpose of the type of the product they wanted to make. Collectively, they agreed they needed a product which could be sold to children for costume dress ups or used for CosPlay (with one sister involved in CosPlay), and saleable as a CraftPlace product for the upcoming Christmas period. Given the product mindset, and desire to derive an income from their prototype, sisters were keen to professionalize their design and ensure product safety. They worked with the DigitalCastle trainer to create a pattern for the crown, and agreed on a set of measures to improve the demo design shown. For example they agreed to add felt at the back, in order to sew the battery in between layers of felt so it could not be accessed by children. They added Velcro for ease of wearing. One sister brought in her own overlocker that she had in her car, to ensure a product level finish. The sisters were all very focused on ensuring they could develop a product that could be sold. They discussed amongst themselves how they could be the trainers, training other sisters to make the crowns, and what they could price them at. One sister stated to the group:

> For me it will be interesting for other sisters. We can start with some workshops like the ones we just had. This is something everyone should know. We can use chiffon, and make a necklace, and a light up flower in the centre. That's good for Christmas. Australia is very lucky to have migrants, we have different talents and skills, and when we sit together, we can come up with ideas that are exceptional!

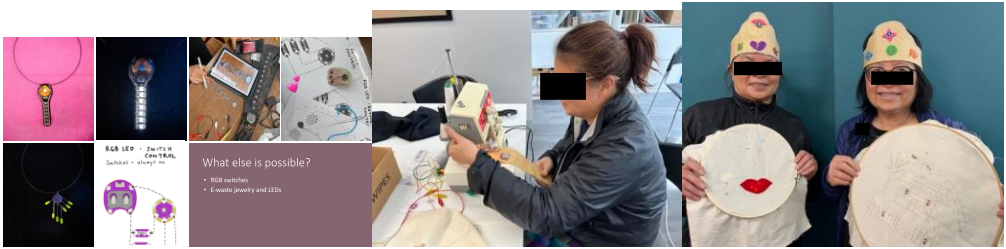

Fig. 5. Day 2 – 1) RGB switches slide 2) sister using her overlocker 3) sisters presenting their work

To conclude the workshop, the sisters were provided with a presentation on circuit playgrounds, and the programming involved to activate the order of lights, sensors and music. They were all incredibly enthusiastic to be given the opportunity to learn this too.

Follow up interviews were scheduled with two of the Group A sisters, during the week after the workshops. Inspired by the workshops, a sister took the initiative to find photos online of what she would like to make and sent them to the first author prior to the interview. When asked what she would like to learn in the future, she said:

> The workshops were good, but I want to go further with this. I want to be able to make light up princess dresses, like the photos I sent you. I can see this as a business in the future because I haven't seen it in Australia. I go to CosPlay, ComicCon, and Lego events every year and I've never seen people with a light up dress. I would like to start with a collaboration with CraftPlace as they can provide how to source good quality fabrics, electronics and I'm not sure if I can afford this myself. They can open gateway on how to deal with supplier, customer. I haven't had a business before, but I can see myself doing this.

The other sister interviewed discussed her motivation to also continue to make e-textiles:

> I actually like the new technology, and now I'm finishing my crown, because my granddaughter wants to have it at the disco. She told me make it strong, sturdy. I can make bible verses with lights on and put it in a frame. Hang it on a wall. "God is Love". I will make something for my granddaughter for her birthday, it's something different for her, it's personalized. I plan to use the skills at home for family and friends. I plan to use it with CraftPlace also with Christmas decorations. We can also make light up scrunchies. In order to sustain, we should make more products, not just what we make now. We can make use of the electronics. Belts with lights on. During parties, in events, I'm always thinking of parties. In the Filipino community, they always have a "Get Connected" day in the Seniors group, and we always have a program where we dress up and make ourselves different. I'm going to make a fascinator that lights up!

## 4.4 The Benefits of Organizational Partnering to Bridge Skills and Diverse Communities (and its Limitations)

While the objective of this study was to bridge WRAMs into STEM skills, a key finding was the value of bridging organizations to diversify skills and participants respectively. As a result of the partnership, CraftPlace gained access to STEM skills and capacity which it had not delivered previously. DigitalCastle gained access to a diversity of participants, WRAMs, which they had not been able to deliver to until these workshops.

The first author was able to broker gains for both partners, in facilitating an ongoing relationship for continuous capacity building for each organization. Given their location in different cities, following the workshops in the joint debriefing interview, they agreed to coach each other in a 'train the trainer' type modality to enable both organizations to continue to grow their capacity and diversity of participants and offerings. For CraftPlace, many questions remained on sourcing equipment and materials and how to structure workshops to continue to offer STEM based activities and building networks to find appropriately skilled trainers. For DigitalCastle, mentorship was required on how to build the cultural capability to create a welcoming environment for WRAMs, how to tap into the right networks, and how to cater to the frequent need to build English skills as part of the training, as well as a sense of belonging, community and social capital.

For both organizations, three factors had potential to impede the continuity of the partnership, and thus continuity of facilitating e-textile trainings and product development with sisters:

- **The viability of having a product to take to market.** While CraftPlace saw the value in offering the training on e-textiles and other STEM skills to sisters for their skill development, in order to sustain their business model, they needed a viable product that could be sold in their retail shop, or an employment pathway for the women with external employers. CraftPlace staff were concerned about any safety issues with making products with button batteries, and any related regulatory issues. The idea of a light-up safety vest for bike riders was discussed for some time, and extended into other ideas such as light up vests for dogs to wear to support night time dog walkers. However, uncertainty about regulatory requirements and product safety remained a concern. The DigitalCastle staff member investigated other electronic materials that could be used, and possible prototype options were discussed, but it was concluded by both organizations that using alternatives would not be financially viable. CraftPlace also concluded they would not be able to afford to order the e-textile parts used at the workshops for the purpose of ongoing training workshops, without it leading to a product. The staff member explained:

> We do get funding from philanthropists and some small grants. It helps cover the gap in our operational costs and what we make from selling products. But I think there are expectations around what refugee and migrant

Building New Clubhouses:-Bridging Refugee and Migrant Women into Technology Design and Production by Leveraging Assets                                                                                                                                                             XX:10women will bring to market and the funding available. We can ask for funding to cover craft materials, sewing machines. We can sell calico bags, jewelry, crocheted items. I don't know how we could get funding to pay for prototyping, patenting, safety testing, electronic parts. It's a lot of risk for us to take on.

- **Employment pathways remained unclear.** While broadening the sisters' skill sets was seen as very positive, all parties remained uncertain as to which employers would value skills in e-textiles, or in other skill areas floated, such as digital fabrication.
- **Geographical distances.** Given the distance between the organizations, it would not be feasible for DigitalCastle to utilize its own budget to continue to train the sisters either, and they would need to find an organization working with refugees and migrants in their hometown.

The organizations agreed to continue to maintain the bridge, and liaise on lessons learned on their respective journeys. However, in practice, the bridge was not maintained and CraftPlace did not continue to offer e-textile or any STEM related workshops to the sisters. Even though the endorsement of the sisters had been secured, the practical realities constrained CraftPlace from further engagement. In the final interview, when the CraftPlace staff member was asked about feedback she received from the sisters, she discussed how enthusiastic the sisters were, but also suggested that CraftPlace would not be supporting further on-site. The staff member said in reference to the sister who wanted to make light up dresses "I think she is happy to go off and buy what she needs. She understands the concept of circuitry now. I think she is planning to make a Cos play outfit for her daughter and herself."

However, when the lead author followed up with this sister nine months after the workshops to see if she had applied her skills, she still wanted to but had not. She said she was disappointed that CraftPlace had not pursued sourcing electronic components or taken any activities further. The cost of sourcing components remained the main constraint to pursue it further independently. DigitalCastle had yet to connect with local organizations supporting refugee and migrant women to involve them in their makerspace and was hoping the first author could help facilitate a connection.

**4.5 Conclusion of Findings**

In coding and analysing the data, it became evident that in taking a community-led design approach with a lived experience, community embedded institution, there was still a clear demarcation between the institution and the community itself as to how they perceived the community's assets and constraints, and whether the resulting dialogue was strengths or deficits based. Table 3 below summarises the assets and constraints of Craftplace as an institution, and that of the WRAM community, as perceived by CraftPlace in Stage 1. The table then reflects the Assets and Constraints of the WRAM Community and CraftPlace as an institution, as perceived by the WRAMs in Stage 2. This is followed by an analysis of the same through the CraftPlace lens in Stage 2.

Table 3: Assets and Constraints

| | Assets | Constraints |
|---|---|---|
| **Stage 1 (As identified by CraftPlace)** | Institutional Assets of CraftPlace:<br>- Safe, trusted venue | WRAM Constraints:<br>- Limited interest in STEM activities including electronics, consciously |

PACM on Human-Computer Interaction, Vol. 5, No. CSCW1, Article XXX, Publication date: April 2021.

| | | |
|---|---|---|
| | - Learning schedule that could 'push' some sisters into attending the workshops<br>- Learning modalities that are effective with WRAMs (eg visual, tactile)<br>WRAM Assets:<br>- Sewing and craft skills<br>- Motivation to make products<br>- Motivation to earn an income | or due to cultural/social constraining factors<br>- May be unwilling to engage due to limited comprehension of the activities<br>- Computer skills are limited and thus any form of programming, even at its most simple, should not be included |
| **Stage 2 (As identified by WRAMs)** | Institutional Assets of CraftPlace:<br>- Safe, trusted, child-friendly venue that fosters a sense of belonging<br>- Learning modalities that are effective when English is constrained (eg visual, tactile)<br><br>WRAM Assets<br>- Growth mindset: Appreciation of new opportunities, appetite to learn new skills including technology<br>- Innovative mindsets: on how to adapt technology to improve usability<br>- Product innovation mindset: for unmet needs of their own communities (in global south and host/new country) and to expand CraftPlace's own product offering<br>- Communal mindset: interest by participants to train other sisters, using a train the trainer model<br>- Goal oriented: wanting to introduce e-textile based products in time for Christmas | WRAM Constraints<br>- Comprehension of English<br>- Not having their skills and experience recognized by Australian employers<br>- Financial, unable to afford materials to continue activities at home<br><br>Institutional Constraints<br>- Lack of interest of CraftPlace to expand product offering |
| **Stage 2 (As identified by CraftPlace)** | Institutional Assets<br>- Staff and volunteer scaffolding of learning<br>- Witnessed and acknowledged sister appetite to engage in e-textiles and new technology | Institutional Constraints:<br>- Financial: resourcing for innovation<br>- Regulatory: concerns on how to take innovative tech-based products to market<br>- Unclear employment pathways for WRAMs with e-textile/tech skills |

While there was a clear overlap in the understanding of the Institutional Assets of CraftPlace between CraftPlace staff and Sisters, there was a tension between Stage 1 and Stage 2 understanding of WRAM assets and appetite to engage in e-textiles, to extend learning, to adapt technology and to innovate.

## 5 DISCUSSION

This paper makes three contributions as outlined in the discussion below. First, we offer a strengths-based counter narrative on the assets, abilities and motivations of WRAMs to engage in makerspaces, particularly STEM skills. Second, we offer a discussion on the implications of racial capitalism and internalized bias which can result in a deficit dialogue that constrains opportunities, limits resources, research and practice with WRAMs and consequently, technological design and production. Third, we extend the work of Buechley and contribute five strategies to bridge WRAM women into STEM oriented makerspace activities to build a "new clubhouse". We discuss the vital role researchers, technologists, makerspaces and financiers must play in supporting these new clubhouses to facilitate strengths-based narratives, harnessing and amplifying skills-based assets, in order to diversify who shapes technology and thus what is shaped.

### 5.1 Counter Narrative on the Motivations, Assets and Abilities of WRAMs

It is well established that makerspaces are meant to democratize access to the means for technological production, and certain equipment and tools available in makerspaces, are meant to lower barriers to entry for those with inexperience in STEM skills [25, 99, 104, 106, 117]. It is also well established that makerspaces struggle to achieve this purpose, and can perpetuate access to those who already dominate STEM fields, particularly white males [6, 26, 74, 79, 104].

What is not well documented is the deficit discourse that can impede attempts at bridging WRAMs into STEM skills via makerspaces. The Stage 1 community-led design process and CraftPlace focus on building from existing assets aligned with best practice [22, 25, 50, 53, 83, 122]. This approach enabled the authors to leverage the venue, learning modalities, environment and scaffolding assets of CraftPlace. In line with Pei's framework, the technology novelty factor was medium, introducing sisters to 'adopting' e-textiles as a method, while leveraging their existing assets in sewing skills, SisterCraft sewing resources, and sister time (using the push method to encourage Group A sisters to stay on from sewing workshops). This assets-based approach lay critically important foundations for the success of the workshops. However, the Stage 2 workshops revealed an unexpected tension regarding the assets-based approach with regards to sister skills and capacity for novelty and technology innovation.

In this instance, a deficit discourse originated with the civil society organization (CSO) that represented the WRAMs, led by staff with lived experience as migrants, albeit not as under-resourced as the sisters themselves. A key lesson that emerged from the sisters in Groups A and B is the need for adoption of a strengths-based counter-narrative, where the planning stages for this research were dominated by a deficit discourse and constraining social constructions of the women's assets, interests and abilities. Even though a best practice community-led design process [22, 53, 93] was instigated and maintained with CraftPlace staff, this was both an enabler and a hindrance as to the range and depth of STEM based activities that could be trialled. This study's findings, particularly in 4.1 and 4.4, presents the beginnings of a counter narrative on the motivations and abilities of WRAMs to engage with STEM skills in makerspaces. While CSCW literature has documented the benefits of strengths-based and assets-based approaches with marginalized communities [83, 122], there is a gap in the strengths-based counter narrative on WRAMs where strengths, assets and capacity for novelty can be underestimated. CSOs may not have the time, capacity or tools to undertake a comprehensive assessment of the intangible and tangible skills-based assets of their own community. Providing tools to CSOs to assess assets may support them to build a counter narrative on the motivations, assets and abilities of their community. Strengths-based counter narratives will help facilitate more partnerships between CSOs and makerspaces and it will support WRAMs who are interested "to see what they can be" through case studies. The women in this study wanted to lead by example, and conduct their own workshops, to demonstrate to other sisters what they could all be capable of, and to innovate together. As stated by one participant in discussing a train the trainer model and products they could produce "Australia is very lucky to have migrants, we have different talents and skills, and when we sit together, we can come up with ideas that are exceptional!" One sister had experience in electronics manufacturing which had previously been unknown to SisterCraft staff. In essence, the sisters wanted to lead a counter narrative to fully articulate their assets, demonstrate their innovative strengths and capacity for novelty. It is contended that without supporting WRAM communities to conduct similar activities and research, the current narrative and reality of severe inequality will persist because there is a risk that assets-based community development approaches underestimate the actual skills-based assets of marginalized communities. Extending the work of Pei [83], while the sustainability of interventions is important, without pursuing high and medium novelty activities, there is a risk that the assets of marginalized communities will continue to be under-estimated. WRAMs and other intersectional groups will remain excluded due to the structural barriers and deficit discourse that dominates, leaving the assets and strengths of these communities untapped.

### 5.2 The Limitations of Resourcing Reflect Racial Capitalism and Stymies Innovation

Makerspaces are a billion dollar industry, with increasing deployment in to public libraries, schools, universities and even in multinational corporations, as well as national and multinational government entities such as the White House, the European Union and more [43, 57, 70, 73, 117]. A key driver behind their popularity is that they spur open innovation, in ways that traditional manufacturing does not enable, leading to unique, bespoke creations [17, 60, 85, 107, 108]. When these innovations are created in makerspaces with the right level of resourcing and social networks, the commercial viability can be tested.

However, not all bespoke creations made in makerspaces have the potential to be commercialized, nor are they necessarily designed with that purpose. CSOs that operate social enterprises, such as CraftPlace, do have makers that are innovative and capable of making products with market appeal, but a CSO type makerspace such as CraftPlace does not have the resources to investigate or seek to ensure regulatory compliance with safety requirements, which stymies innovation. The lack of motivation of CraftPlace to continue STEM based activities came down to resourcing, with concerns on funding and creating a viable and legally compliant commercial product. The sisters were not discussing anything too sophisticated, but rather light up costumes and jewellery, or light up vests for humans and dogs. Rather than innovation, adoption or amplification, in terms of Pei's novelty scale [83], the sisters were seeking to adopt and then adapt technology to suit their unique needs and leverage their strengths, representing an extension of novelty beyond technology adoption. However, the organization that was best placed to enable the women to pursue technology adoption and adaption felt constrained by explicit financial and legal barriers. It is herewith contended that implicit structural barriers in the form of systemic racial capitalism [113] constrained CraftPlace to pursue medium or higher novelty activities, rather than the skills based assets of the sisters themselves. The staff member's own views were shaped by her interactions with philanthropists and grant makers, and what she perceives they are willing to fund. She said "I think there are expectations around what refugee and migrant women will bring to market…". Although this was just one statement offered by the staff member at the concluding interview, the actions of CraftPlace to not take this work forward despite the sister demand paints a broader picture that is in stark contrast to the billion-dollar makerspace industry. CSOs, even when led by members of the community with lived experience, may be in a pattern, consciously or subconsciously, of working with funders, partners and researchers that require a needs-based and deficits-based dialogue with regards to the capacity of their community. This deficits-based dialogue may be essential to obtain and sustain resourcing and programming. Being in a pattern of working with this deficit lens can result in a limitation of understanding of skills-based assets, which would enable a more comprehensive strengths-based approach.

As CSCW researchers, we need to interrogate the hidden biases on what WRAMs in a western context produce (ie gendered craft-oriented products [54, 114]) and deficit discourses around skills, even when the discourse originates at CSOs and are embedded within community-led, assets based approach. Turner discusses the need for creative professionals of color to perform race [113], resulting in products and services that fit mainstream community expectations and biases, rather than being authentic to the creative's assets and capacity for expression. Nakamura also discusses the role of women of colour in digital labour, and how these women have been exploited in manufacturing electronic circuits through racialised, gendered dialogues regarding women of colour's dexterous fingers and how circuit work aligns with traditional craft practices [77]. Knowles examines how older adults are limited by dialogue around accessibility and limitations of age, rather than interrogating the skills and capacity older adults can actually bring [69]. We acknowledge that we are speculating, but it is possible that CraftPlace staff feel obliged to produce products that perform "female refugee", due to philanthropic biases and how they perceive market demand. We also acknowledge that this very study risks replicating the racialised, gendered narrative Nakamura was questioning. Questions around the lens applied to how we interpret and value female women of color interests, and capabilities with regards to making have been questioned by Vossoughi [114], but the resourcing and systemic capitalist implications remain under-researched. There is a role for CSOs and makerspaces to disrupt exploitative practices and support genuine assets-based interventions that foster technology design and production. However, this could rely on researchers providing the methodology and tools to conduct a comprehensive audit of skills-based assets that may have remained unseen and untested. As discussed above, CraftPlace staff had their own internalized biases regarding the sister's interests, assets and capabilities, but also what funders would be interested in. Their assumptions that philanthropists or governments would not provide funding to explore STEM skill development and STEM oriented craft products remains untested. Further research needs to be conducted on how racial capitalism impedes diversification of the makerspace movement and its innovations. There is also potential for a counter narrative on funder motivations to be proffered in this space as well.

One of the findings from the study was the insights the sisters offered in wanting to not just work with the technology to make something, but to have the ability to modify the technology itself for better design. To reframe this in the context of Pei's framework, the sisters wanted to not only engage in medium novelty activities of adopting new technologies in their community, but also in high novelty activities such as adapting technologies. Technology adaption is a factor that is missing from Pei's framework and important to marginalised communities, to draw on their assets and adapt technologies to their particular needs in order to strengthen their asset base. Ideas on changing the shape of the LED sequins and having them machine sewable show the innovation of this group. Turakhia and her contemporaries argue makerspace tools and systems can be improved by centring design around the learner, instructor and learning process, rather than the technology itself [111]. Based on this research, we agree with this proposition, and that this will facilitate a more inclusive approach to technological design and production, with a much broader range of potential innovations. While innovations such as the LilyPad Arduino have benefitted large audiences of women, it remains unclear to those on the ground how they can make their own versions of e-textile tools in line with their own visions and ideas for improvement. Research and sharing of knowledge on how marginalized communities can work with engineering schools and companies to design and produce their own components is required. Further, researchers and technologists need to consider what role they can play in fostering innovation at the social enterprise level, for localized rather than necessarily scalable consumption. However legal compliance and affordability remains essential. Further bridges need to be built to make it possible to ensure craft-based refugee and migrant organizations can equip their makerspaces with customised materials and tools of production that can be utilised in saleable products, to make their innovations viable.

On the other hand, it is also noted that one sister thought about applications for light up clothing to support the safety of people walking at night in non-electrified locations in her country of origin. This is an issue that does need a scalable solution, and there is of course plenty of potential for WRAMs to innovate and lead large scale businesses with international markets.

If resourcing could be secured for organizations such as CraftPlace to support WRAMs to engage in STEM skill training and product development, and associated costs such as legal compliance for safety of products containing electronics, this could be a stepping-stone towards a broader movement of diversifying who is involved in technological design and production be at small, medium or large scale. This disruption is urgently needed at all market entry points.

Finally, a further hindrance was encountered to maintain the enthusiasm of CraftPlace to offer STEM skills training to their sisters. Low hanging fruit of gendered skills in sewing could lead to casual employment with external employers in the garment industry. Beyond soft skills, it was not evident to the research team and partner organizations what linkages could be fostered to employers on completion of e-textile training, or other basic STEM skills as offered in makerspaces. While pathways may be evident to those in the existing clubhouses, those on the outside are not clear and bridges need to be built to enable new clubhouses to support these pathways. While there is some research on entrepreneurship [61], there is a lack of research into employment pathways from makerspaces. This is needed to help foster a diversity of talent to enter into new sectors, for people who are not in a position to follow formal education pathways.

**5.3 Five Strategies to Bridge WRAMs into Technology Design and Production**

Buechley's work to bridge women from craft to electronics laid the foundations for this research [13]. The authors of this paper were unable to locate any literature documenting research into leveraging e-textile work and tools specifically with women of color to leverage their entry into STEM skills. Buechley has stated "instead of trying to fit people into existing engineering cultures, it may be more constructive to try to spark and support new cultures, to build new clubhouses"[13]. Setting up a new makerspace specifically for the study would carry substantial risk of lack of interest from WRAMs due to lack of relationships and trust. Thus, the authors thought it best to follow the best practices of community-led design [22, 53, 71, 93], and center an existing CSO embedded in the WRAM community as the clubhouse, sparking and supporting a new STEM culture at their clubhouse. This study contributes to Buechley's pioneering research with an intersectional lens. It outlines the additional constraints faced by women of color, in this case WRAMs, and the additional bridges that were required to enable their engagement beyond the e-textile activities and tools. Without these bridges, the WRAMs would not have been able to enter the new STEM oriented clubhouse and their assets and strengths would remain untapped. In terms of Buechley's statement "it may be more constructive to try to spark and support new cultures", we recommend 5 strategies across this paper to bridge WRAMs (and people of intersectional identities more broadly) into technology design and production, as summarized here:

1) Facilitate strategic partnerships between community-led CSOs that are embedded in and work closely with marginalized groups and makerspaces that bring the technical STEM skills. Strategically, it makes most sense to partner CSOs that already engage in making of some kind, such as those aimed at entrepreneurship. It also makes most sense to find makerspaces that are more oriented towards empowering marginalized communities, such as a feminist hackerspace.

2) Bridge STEM skills into a culturally safe and tailored learning environment for WRAMs with limited literacy and education in the host country, by
    a. elevating visual content for explanations and inspiration, including photos and videos. The content producing in strategy 1 above could also be potentially utilized to diversify available images and videos of women of color engaging in e-textiles and STEM oriented maker activities
    b. offering tactile opportunities to engage with the materials and equipment.
    c. leveraging staff, community volunteers and peers to enable comprehension.

3) Conduct further research and practice to build a strengths-based counter narrative on the motivations, assets and abilities of WRAMs (and people with other intersectional identities) to engage with and innovate in STEM activities, disrupting the deficit discourse. This may require active interventions by researchers to provide opportunities for participants to reveal assets that are often not perceived or obvious in traditional community development activities run by CSOs. It could involve confronting internalized biases by the community leaders and communities themselves, as well as the researcher's own biases. Assets may be revealed through dialogue and interview, but it may also require autodidactic activities, and opportunities to 'play' in an innovation sandpit, taking a graduated process to increase skill/complexity levels while also allowing freedom to adapt the intervention and technologies based on participant and community assets and interests.

4) Facilitate strategic partnerships between relevant CSOs and researchers, technologists and philanthropic finance to disrupt racial capitalism and catalyze innovation, including of commercially and legally viable products and appropriate market entry by WRAMs. This may involve technology adaption for marginalized community contexts as well as innovation. In pursuing this bridge, there is also potential for a counter narrative on funder and industry motivations to support WRAM innovation to be proffered in this space as well.

5) Conduct further research on pathways from makerspace learning environments into employment in STEM oriented industries, mindful of the histories of exploitation of women of colour in electronics manufacturing [77], and temptations to creative a false empowerment narrative, rather than a genuine empowerment narrative.

These bridges could lead to transformational change, enabling more people from disadvantaged intersectional backgrounds to move into technology design and production, while also transforming existing engineering cultures through cultural transference in both directions.

## 5.4 Limitations

While this research aimed to build bridges, the structures did not hold. As is evident, this was just one small study, with actual STEM oriented maker activity with refugee, asylum seeker and migrant women conducted over only 2 days and confined to e-textiles. Logistical and resourcing mechanisms constrained the completion and sustainability of the bridges. While we can proffer foundational lessons, there is a vital need to extend this research, and these bridging activities on a continuous basis.

## 6 CONCLUSION

Makerspaces hold a utopian vision of disrupting who shapes technology and what is shaped. Makerspaces and HCI researchers alone cannot bridge the divide in who designs technology, but they are vital actors in a leading role. CSOs that serve refugee, asylum seeker and migrant women also can't build a new clubhouse on their own. Partnerships, collaboration and further research is needed, to build the bridges and thus capacity of CSOs to offer leapfrog opportunities for those who have always missed out. Without the current actors in the field building these bridges to a new clubhouse to engage those at the intersections, makerspaces and HCI researchers will continue to reproduce patriarchal structures and not realize their vision.

## ACKNOWLEDGMENTS